\documentclass{ws-procs9x6}
\usepackage{amssymb,amsmath,amsfonts}
\usepackage{graphicx}
\usepackage{graphics}
\usepackage{eepic,epsfig}

\begin{document}

\title{TOPOLOGICAL CASIMIR EFFECT IN \\
POWER-LAW FRW COSMOLOGIES}
\author{A. L. MKHITARYAN and A. A. SAHARIAN$^*$}

\address{Department of Physics, Yerevan State University,\\
1 Alex Manoogian Street, 0025 Yerevan, Armenia\\
$^*$E-mail: saharian@ictp.it}

\begin{abstract}
We investigate the vacuum expectation values of the field squared and the
energy-momentum tensor for a massless scalar field with general curvature
coupling parameter in spatially flat Friedmann-Robertson-Walker universes
with an arbitrary number of toroidally compactified dimensions. When the
comoving lengths of the compact dimensions are short compared to the Hubble
length, the topological parts coincide with those for a conformal coupling.
This limit corresponds to the adiabatic approximation. In the opposite limit
of large comoving lengths of the compact dimensions, in dependence of the
curvature coupling parameter, two regimes are realized with monotonic or
oscillatory behavior of the vacuum expectation values.
\end{abstract}


\bodymatter

\section{Introduction}

In the present talk, based on Ref.~\cite{Saha09}, we consider an
exactly soluble problem for the topological Casimir effect on
background of Friedmann-Robertson-Walker (FRW) universes with a
power-law scale factor. The vacuum polarization and the particle
creation in the FRW cosmological models with trivial topology have
been considered in a large number of papers (see
Refs.~\cite{Birr82,Most97}). In particular, the vacuum expectation
values of the field squared and the energy-momentum tensor in
models with power law scale factors have been discussed in
Refs.~\cite{Bord97}.

In most work on the topological Casimir effect in cosmological
backgrounds, the results for the corresponding static counterparts
were used replacing the static length scales by comoving lengths
in the cosmological bulk. This procedure is valid in conformally
invariant situations or under the assumption of a quasi-adiabatic
approximation. For non-conformal fields the calculations should be
done directly within the framework of quantum field theory on
time-dependent backgrounds (for the topological Casimir effect in
toroidally compactified de Sitter spacetime see Refs.
\cite{Saha08}).

The paper is organized as follows. In the next section we evaluate the
vacuum expectation value (VEV) of the field squared in spatially flat FRW
model with topology $R^{p}\times (S^{1})^{q}$. In Sec. \ref{sec:EMT} we
consider the VEV of the energy-momentum tensor. The main results are
summarized in Sec. \ref{sec:Conc}.

\section{VEV of the field squared}

\label{sec:vevphi2}

We consider a scalar field with curvature coupling parameter $\xi $\
evolving on background of the $(D+1)$-dimensional spatially flat FRW
spacetime with power law scale factor $a(t)=\alpha t^{c}$. In addition to
the synchronous time coordinate $t$ it is convenient to introduce the
conformal time $\tau $ in accordance with $t=[\alpha (1-c)\tau ]^{1/(1-c)}$.
Here we assume that $c\neq 1$. Note that one has $0\leqslant \tau <\infty $
for $0<c<1$ and $-\infty <\tau \leqslant 0$ for $c>1$.

We will assume that the spatial coordinates $z^{l}$, $l=p+1,\ldots ,D$, are
compactified to $S^{1}$: $0\leqslant z^{l}\leqslant L_{l}$, and for the
other coordinates we have $-\infty < z^{l}< +\infty $, $l=1,\ldots ,p$.
Hence, we consider the spatial topology $R^{p}\times (S^{1})^{q}$ with $%
p+q=D $. Along the compact dimensions we will consider the boundary
conditions $\varphi (\tau ,\mathbf{z}_{p},\mathbf{z}_{q}+\mathbf{e}%
_{l}L_{l})=e^{2\pi i\alpha _{l}}\varphi (\tau ,\mathbf{z}_{p},\mathbf{z}_{q})
$ with constant phases $\alpha _{l}$, where $\mathbf{z}_{p}=(z^{1},\ldots
,z^{p})$, $\mathbf{z}_{q}=(z^{p+1},\ldots ,z^{D})$, and $\mathbf{e}_{l}$, $%
l=p+1,\ldots ,D$, is the unit vector along the direction $z^{l}$.

For the VEV of the field squared we have the following
decomposition:
\begin{equation}
\langle \varphi ^{2}\rangle _{p,q}=\langle \varphi ^{2}\rangle _{\mathrm{FRW}%
}+\langle \varphi ^{2}\rangle _{p,q}^{\mathrm{(t)}},\;\langle \varphi
^{2}\rangle _{p,q}^{\mathrm{(t)}}=\sum_{j=p}^{D-1}\Delta _{j+1}\langle
\varphi ^{2}\rangle _{j,D-j},  \label{phi2decomp2}
\end{equation}%
where $\langle \varphi ^{2}\rangle _{\mathrm{FRW}}=\langle \varphi
^{2}\rangle _{D,0}$ is the VEV in the spatial topology $R^{D}$ and the part $%
\langle \varphi ^{2}\rangle _{p,q}^{\mathrm{(t)}}$ is induced by
the nontrivial topology. For the topological part induced due to
the compactness of the $z^{p+1}$ - direction we have
\begin{eqnarray}
&&\Delta _{p+1}\langle \varphi ^{2}\rangle _{p,q}=\frac{4A\eta ^{2b}}{(2\pi
)^{(p+3)/2}V_{q-1}}\sum_{\mathbf{n}_{q-1}\in \mathbf{Z}^{q-1}}\int_{0}^{%
\infty }dyy\left[ I_{-\nu }(y\eta )+I_{\nu }(y\eta )\right]  \notag \\
&&\times K_{\nu }(y\eta ) \sum_{n=1}^{\infty }\frac{\cos (2\pi n\alpha _{p+1})%
}{(nL_{p+1})^{p-1}}f_{(p-1)/2}(nL_{p+1}\sqrt{y^{2}+\mathbf{k}_{\mathbf{n}%
_{q}-1}^{2}}),  \label{phi2}
\end{eqnarray}%
where $I_{\nu }(z)$ and $K_{\nu }(z)$ are the modified Bessel
functions, $f_{\nu }(x)=x^{\nu }K_{\nu }(x)$, and $A=\alpha
^{1-D}[\alpha |1-c|]^{(D-1)c/(c-1)}$. Here the notations $\eta
=|\tau |$, $b=(cD-1)/[2(c-1)]$, $V_{q-1}=L_{p+2},\ldots ,L_{D}$ and $\mathbf{%
n}_{q-1}=(n_{p+2},\ldots ,n_{D})$, $\mathbf{k}_{\mathbf{n}%
_{q-1}}^{2}=\sum_{l=p+2}^{D}(2\pi n_{l}/L_{l})^{2}$, are
introduced. The parameter $\nu $ is defined as
\begin{equation}
\nu =\frac{1}{2|1-c|}\sqrt{(cD-1)^{2}-4\xi Dc\left[ (D+1)c-2\right] }.
\label{nu}
\end{equation}

\begin{figure}[tbph]
\begin{center}
\psfig{figure=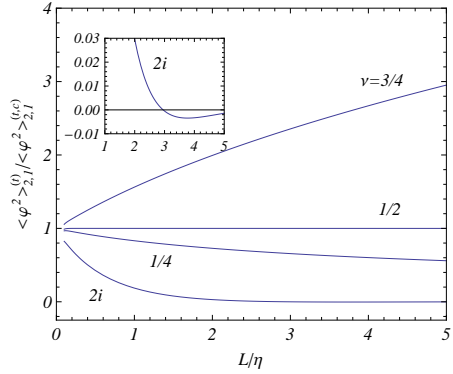,width=5.88cm,height=5cm}
\end{center}
\caption{The ratio $\langle \protect\varphi ^{2}\rangle _{D-1,1}^{\mathrm{(t)%
}}/\langle \protect\varphi ^{2}\rangle _{D-1,1}^{\mathrm{(t,c)}}$ versus $L/%
\protect\eta $ for different values of the parameter $\protect\nu $ (numbers
near the curves) and for $D=3$ scalar field with the periodicity condition
along the compact dimension.}
\label{fig1}
\end{figure}

In figure \fref{fig1}, we have plotted the ratio $\langle \varphi
^{2}\rangle _{D-1,1}^{\mathrm{(t)}}/\langle \varphi ^{2}\rangle _{D-1,1}^{%
\mathrm{(t,c)}}$ for the special case of topology $R^{D-1}\times S^{1}$ as a
function of $L/\eta $, with $L=L_{D}$ being the length of the compact
dimension, for untwisted $D=3$ scalar field ($\alpha _{D}=0$) and for
various values of the parameter $\nu $. Note that the ratio $L/\eta $ is
related to the comoving length of the compact dimension, measured in units
of the Hubble length, by $L_{l}/\eta =(|1-c|/c)L_{l}^{\mathrm{(c)}}/r_{%
\mathrm{H}}$. Figure \ref{fig1} clearly shows that the adiabatic
approximation for the topological part is valid only for small
values of the ratio $L^{\mathrm{(c)}}/r_{\mathrm{H}}$.

In the limit $L_{l}^{\mathrm{(c)}}\ll r_{\mathrm{H}}$ the
topological part coincides with that for a conformal coupling and
behaves like $\langle \varphi ^{2}\rangle
_{p,q}^{\mathrm{(t)}}\propto t^{c(1-D)}$. This limit corresponds
to the adiabatic approximation. In the opposite limit the behavior
of the VEV is qualitatively different for real and imaginary
values of the parameter $\nu $. For real values, the topological
part behaves as $\langle \varphi ^{2}\rangle
_{p,q}^{\mathrm{(t)}}\propto t^{2(c-1)\nu -cD+1}$. In the limit
$L_{l}^{\mathrm{(c)}}\gg r_{\mathrm{H}}$
and for imaginary values $\nu $ the asymptotic behavior is oscillatory: $%
\langle \varphi ^{2}\rangle _{p,q}^{\mathrm{(t)}}\propto
t^{1-cD}\cos [2|\nu |(c-1)\ln (t/t_{0})+\psi ]$.

\section{VEV of the energy-momentum tensor}

\label{sec:EMT}

For the the VEV of the energy-momentum tensor we have the formula
\begin{equation}
\langle T_{i}^{k}\rangle _{p,q}=\langle T_{i}^{k}\rangle _{\mathrm{FRW}%
}+\langle T_{i}^{k}\rangle _{p,q}^{\mathrm{(t)}},\;\langle T_{i}^{k}\rangle
_{p,q}^{\mathrm{(t)}}=\sum_{j=p}^{D-1}\Delta _{j+1}\langle T_{i}^{k}\rangle
_{j,D-j},  \label{TikDecomp}
\end{equation}%
where $\langle T_{i}^{k}\rangle _{\mathrm{FRW}}$ is the part corresponding
to the uncompactified FRW spacetime and $\langle T_{i}^{k}\rangle _{p,q}^{%
\mathrm{(t)}}$ is induced by the the nontrivial topology. The first term is
well investigated in the literature. Here for the topological part we have
(no summation over $i$)%
\begin{eqnarray}
&&\Delta _{p+1}\langle T_{i}^{i}\rangle _{p,q}=\frac{4A\Omega ^{-2}}{(2\pi
)^{(p+3)/2}V_{q-1}}\sum_{\mathbf{n}_{q-1}\in \mathbf{Z}^{q-1}}\sum_{n=1}^{%
\infty }\frac{\cos (2\pi n\alpha _{p+1})}{(nL_{p+1})^{p-1}}\int_{0}^{\infty
}dy\,y^{3-2b}  \notag \\
&&\times \left[ f_{(p-1)/2}(z)F^{(i)}(\eta y)-f_{p}^{(i)}(z)\frac{\tilde{%
I}_{\nu }(\eta y)\tilde{K}_{\nu }(\eta y)}{(nL_{p+1}y)^{2}}\right]
_{z=nL_{p+1}\sqrt{y^{2}+\mathbf{k}_{\mathbf{n}_{q}-1}^{2}}},
\label{Tiin}
\end{eqnarray}
with $\tilde{K}_{\nu }(z)=z^{b}K_{\nu }(z)$, $\tilde{I}%
_{\nu }(z)=z^{b}\left[ I_{\nu }(z)+I_{-\nu }(z)\right] $, and
\begin{eqnarray}
F^{(0)}(z) &=&\frac{1}{2}\tilde{I}_{\nu }^{\prime }\tilde{K}_{\nu
}^{\prime }+\frac{D\xi c}{z(1-c)}(\tilde{I}_{\nu }\tilde{K}_{\nu
})^{\prime }-\frac{1}{2}\left[ 1-\frac{\xi D(D-1)c^{2}}{z^{2}(1-c)^{2}}%
\right] \tilde{I}_{\nu }\tilde{K}_{\nu },  \notag \\
F^{(l)}(z) &=&2\left( \xi -\frac{1}{4}\right) \tilde{I}_{\nu }^{\prime }%
\tilde{K}_{\nu }^{\prime }-\frac{1}{z}\frac{c\xi }{1-c}(\tilde{I}%
_{\nu }\tilde{K}_{\nu })^{\prime }  \label{F(z)} \\
&&+2\left[ \xi -\frac{1}{4}-Dc\xi \frac{(Dc-2)(\xi -\xi _{D})+\xi c}{%
z^{2}(1-c)^{2}}\right] \tilde{I}_{\nu }\tilde{K}_{\nu }, \notag
\end{eqnarray}%
where $l=1,\ldots ,D$. In Eq. (\ref{Tiin}) we have used the
notations
\begin{eqnarray}
&& f_{p}^{(0)}(z)
=0,\;f_{p}^{(i)}(z)=f_{(p+1)/2}(z),%
\;f_{p}^{(l)}(z)=(nL_{p+1}k_{l})^{2}f_{(p-1)/2}(z),  \notag \\
&& f_{p}^{(p+1)}(z) =-pf_{(p+1)/2}(z)-z^{2}f_{(p-1)/2}(z),
\label{fz}
\end{eqnarray}%
where $i=1,2,\ldots p$ and $l=p+2,\ldots D$.

For $L_{l}^{\mathrm{(c)}}\ll r_{\mathrm{H}}$ the topological part
behaves like $\langle T_{i}^{k}\rangle
_{p,q}^{\mathrm{(t)}}\propto t^{-c(D+1)}$ and to the leading order
the stresses along the uncompactified dimensions are equal to the
vacuum energy density. In the limit $L_{l}^{\mathrm{(c)}}\gg
r_{\mathrm{H}}$ and for real values $\nu $ the asymptotic has the
form $\langle T_{i}^{k}\rangle _{p,q}^{\mathrm{(t)}}\propto
t^{2(c-1)\nu -cD-1}$. The corresponding vacuum stresses are
isotropic and the equation of state for the topological parts
in the vacuum energy density and pressures is of the barotropic type. For $%
L_{l}^{\mathrm{(c)}}\gg r_{\mathrm{H}}$ and for imaginary values $\nu $ we
have the asymptotic behavior $\langle T_{i}^{k}\rangle _{p,q}^{\mathrm{(t)}%
}\propto t^{-cD-1}\cos [2|\nu |(c-1)\ln (t/t_{0})+\psi _{i}]$.

\section{Conclusion}

\label{sec:Conc}

We have investigated one-loop quantum effects for a scalar field with
general curvature coupling, induced by toroidal compactification of spatial
dimensions in spatially flat FRW cosmological models with power law scale
factor. General boundary conditions with arbitrary phases are considered
along compact dimensions. The boundary conditions imposed on possible field
configurations change the spectrum of vacuum fluctuations. Among the most
important characteristics of the vacuum state are the expectation values of
the field squared and the energy-momentum tensor. Though the corresponding
operators are local, due to the global nature of the vacuum these VEVs carry
an important information on the global structure of the background
spacetime. We present the VEVs as the sum of the function for topologically
trivial FRW model and the topological part. The latter is finite in the
coincidence limit and in this way the renormalization of the VEVs is reduced
to that for the FRW universe with trivial topology. The topological parts
are given by formulae (\ref{phi2decomp2}) and (\ref{phi2}) for the field
squared and by formulae (\ref{TikDecomp}), (\ref{Tiin}) for the energy
density and the stresses. A.L.M. gratefully acknowledges the organizers of
the conference QFEXT09 for the opportunity to present this paper.

\bibliographystyle{ws-procs9x6}
\bibliography{ws-pro-sample}

\begin{thebibliography}{9}
\bibitem{Saha09} A. A. Saharian and A. L. Mkhitaryan, arXiv:0908.3291.

\bibitem{Birr82} N. D. Birrel and P. C. W. Davies, \textit{Quantum Fields in
Curved Space} (Cambridge University Press, Cambridge, 1982); A. A. Grib, S.
G. Mamayev, and V. M. Mostepanenko, \textit{Vacuum Quantum Effects in Strong
Fields} (Friedmann Laboratory Publishing, St. Petersburg, 1994).

\bibitem{Most97} V. M. Mostepanenko and N.N. Trunov, \textit{The Casimir
Effect and Its Applications} (Clarendon, Oxford, 1997); K. A.
Milton, \textit{The Casimir Effect: Physical Manifestation of
Zero-Point Energy} (World Scientific, Singapore, 2002); M. Bordag,
G. L. Klimchitskaya, U. Mohideen and V. M. Mostepanenko,
\textit{Advances in the Casimir Effect} (Oxford University Press,
Oxford, 2009).

\bibitem{Bord97} M. Bordag, J. Lindig, V. M. Mostepanenko, and Yu. V. Pavlov,
\textit{Int. J. Mod. Phys. D} \textbf{6}, 449 (1997); M. Bordag,
J. Lindig, and V. M. Mostepanenko, \textit{Class. Quantum Grav.}
\textbf{15}, 581 (1998).

\bibitem{Saha08} A. A. Saharian and M. R. Setare, Phys. Lett. B \textbf{659},
367 (2008); S. Bellucci and A. A. Saharian, Phys. Rev. D
\textbf{77}, 124010 (2008); A. A. Saharian, Class. Quantum Grav.
\textbf{25}, 165012 (2008); E. R. Bezerra de Mello and A. A.
Saharian, JHEP \textbf{12}, 081 (2008).
\end{thebibliography}

\end{document}